\documentclass[fleqn,usenatbib]{mnras}

\usepackage[T1]{fontenc}

\usepackage{graphicx}   % Including figure files
\usepackage{amsmath}    % Advanced maths commands
\usepackage{amssymb}    % Extra maths symbols
\usepackage{subfigure}
\usepackage{makecell}
\usepackage{threeparttable}
\usepackage{supertabular}
\usepackage{longtable}

\title[Terrestrial planet formation around M Dwarfs]{The Terrestrial Planet Formation around M Dwarfs: In-situ, Inward Migration or Reversed Migration}

\author[Pan Mengrui et al.]{
Mengrui Pan, $^{1,2}$
Su Wang, $^{1,2}$\thanks{}
Jianghui Ji$^{1,2,3}$\thanks{E-mail: wangsu@pmo.ac.cn, jijh@pmo.ac.cn}
\\
$^{1}$CAS Key Laboratory of Planetary Sciences, Purple Mountain Observatory, Chinese Academy of Sciences, Nanjing 210023, China\\
$^{2}$School of Astronomy and Space Science, University of Science and Technology of China, Hefei 230026, China\\
$^{3}$CAS Center for Excellence in Comparative Planetology, Hefei 230026, China\\
}

\date{Accepted XXX. Received YYY; in original form ZZZ}

\pubyear{2021}

\begin{document}
\label{firstpage}
\pagerange{\pageref{firstpage}--\pageref{lastpage}}
\maketitle

% Abstract should update.
\begin{abstract}
Terrestrial planets are commonly observed to orbit M dwarfs with close-in trajectories.
In this work, we extensively perform N-body simulations of planetesimal accretion with three models of in-situ, inward migration and reversed migration to explore terrestrial formation in tightly compact systems of M dwarfs.
In the simulations, the solid disks are assumed to be 0.01\% of the masses of host stars and spread from 0.01 to 0.5 AU with the surface density profile scaling with $r^{-k}$ according to the observations.
Our results show that in-situ scenario may produce $7.77^{+3.23}_{-3.77}$ terrestrial planets with an average mass of $1.23^{+4.01}_{-0.93} \ M_{\oplus}$ around M dwarfs.
The number of planets tends to increase as the disk slope is steeper or with a larger stellar mass.
Moreover, we show that $2.55^{+1.45}_{-1.55}$ planets with mass of $3.76^{+8.77}_{-3.46} \ M_{\oplus}$ are formed in the systems via inward migration, while $2.85^{+1.15}_{-0.85}$ planets with $3.01^{+13.77}_{-2.71} \ M_{\oplus}$ are yielded under reversed migration.
Migration scenarios can also deliver plentiful water from the exterior of ice line to the interior due to more efficient accretion.
The simulation outcomes of reversed migration model produce the best matching with observations, being suggestive of a likely mechanism for planetary formation around M dwarfs.
\end{abstract}

\section{Introduction}

\begin{figure}
\includegraphics[width=\columnwidth]{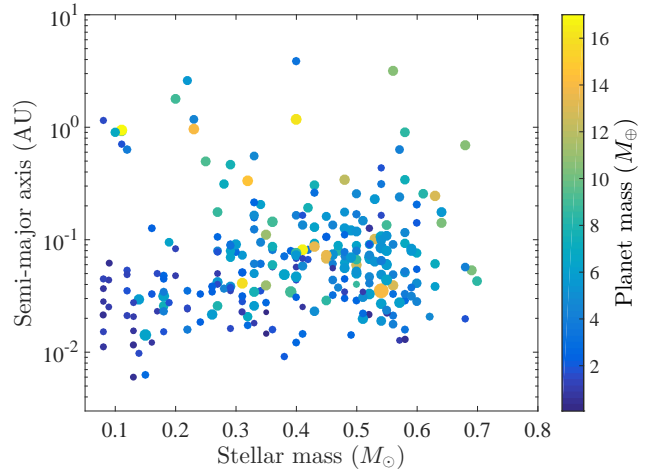}
\caption{The distribution of planets around M dwarfs. The color and size of filled circles are scaling as the radii and masses of planets, respectively.
%in which red and green dots indicate planets whose masses under 10 $M_{\oplus}$ and 10-100 $M_{\oplus}$  orbiting M Dwarfs, respectively.
%The blue circles are giant planets with masses above 100 $M_{\oplus}$ and grey dots are planets with unknown masses.
\label{fig:a_M}}
\end{figure}

M dwarfs, which are the most abundant main-sequence stars in the Milky Way, host about 40\% masses of our galaxy and play a crucial role in revealing the formation and evolution of stars and planets \citep{Chabrier2003, Johnson2007}.
As of nowadays, nearly 4800 exoplanets have been discovered, among which $\sim$ 10\% orbit M dwarfs \citep{Bonfils2013, Winters2015}.
Planets around M dwarfs are generally observed in the high-levels of X-ray and ultraviolet radiation because of the stellar magnetic activity \citep{Segura2010}, leading to huge variations in the planets around M dwarf and other main-sequence stars.
In addition, the luminosity of M dwarfs is considerably low that searching for transiting planets in habitable zone (HZ) are much easier than around FGK stars \citep{Nowak2020, Luque2021}.
Figure \ref{fig:a_M} shows the orbital distribution of 273 terrestrial planets or sub-Neptune with masses smaller than 17 $M_{\oplus}$ around M dwarfs or stars with masses between 0.075 and 0.6 $M_{\odot}$ \citep{Mann2015}.
The provided census further suggests that nearly 89\% of planets locate at 0.01 to 0.3 AU away from their host stars, which is more likely to be detected combined with high occurrence rate of terrestrial planets around M dwarfs \citep{Fressin2013, Silburt2015, Mulders2015, Hsu2020}.  Therefore, the habitability, occurrence, formation and evolution of terrestrial planets around M dwarfs are of particular interests to the community.

Several scenarios, such as in-situ formation, inward migration, or pebble-accretion, have been proposed to elucidate terrestrial planet formation around M dwarfs \citep{Bond2010, Ogihara2015, Lee2017, Unterborn2018, Bitsch2019}.
\citet{Raymond2007} performed N-body simulations to explore in-situ accretion of the planetary embryos around low-mass stars and indicated that few planets around M stars could grow up to sufficient mass (> 0.3 $M_{\oplus}$) and be habitable in the minimum-mass solar nebula (MMSN) model.
However, remarkably massive solid disk ($\sim$ 50-100 $M_{\oplus}$ interior to 1 AU) is still required to reproduce the observed distribution of hot Neptune and super-Earth \citep{Hansen2012}.
As an alternative formation theory, by taking tidal interaction with gaseous disk into account, \citet{Ogihara2009} emphasized that type I migration rate plays a key part in reshaping final configuration of terrestrial planets. After efficient type I migration, 4 to 6 Earth-like planets are eventually formed through planetesimal accretion, whereas $\sim$ 40 small protoplanets still survive in gaseous disk via slow migration \citep{Ogihara2009}.
However, the migration of planets may be reversed at the inner region of gaseous disk, as a result of the transition of heating mechanism from viscously heated to the radiatively heated in the self-consistent non-isothermal disk \citep{Garaud2007}.
Compact planetary systems that harbor one to seven planets with high water percentage could form commonly posterior to reversed migration around low-mass stars and brown dwarfs \citep{Miguel2020}.
Furthermore, inside-out planetary formation model indicates that planets in the inner disk are produced sequentially at the pressure maximum by pebble collision or near snow line by streaming instabilities, thereby revealing formation of tightly-packed planetary system \citep{Chatterjee2014, Ormel2017}.
Besides, tidal circularization, secular resonance sweeping, planetary interaction as well as stellar magnetosphere may also rebuild planetary  configurations to yield compact and close-in systems \citep{Nagasawa2011, Pan2020, Liu2020b}.

The aforementioned mechanisms are summarized as in-situ collision, inward migration and reversed migration, among which inward migration is the most widely accepted scenario even though several conundrums remain unsolved, including the birthplace of protoplanet and the timescale of type I migration \citep{Fortier2013, Johansen2019}.
Here we adopt three kinds of mechanisms to investigate terrestrial planet formation under the circumstances that the disks are with a variety of surface density profiles in combination with M dwarfs of diverse stellar masses \citep{Bolmont2014}.
Compared to the investigation of the growth of less massive embryos in large scale disk \citep{Alibert2017, Miguel2020}, we conduct the study to explore the late-stage scenario of planetary formation in the inner region of disks (0.01-0.5 AU) with dozens of embryos that have accreted efficiently.
The simulation results are expected to reproduce the observations and provide a better understanding of terrestrial planet formation around M dwarfs.
For in-situ formation, we assume the gaseous disk surrounding host stars is dissipated \citep{Bolmont2014}.
Only embryos and planetesimals embedded in solid disk interact each other via gravity.
Tidal interaction with isothermal and non-isothermal gaseous disk is considered under the scenarios of inward and reversed migration, respectively.

This paper is structured as follows. In Section \ref{sec:model}, we describe the disk models and characteristic timescales for each scenario.
In Section \ref{sec:simulation}, we extensively show simulation outcomes for terrestrial planet formation around M Dwarfs via in-situ, inward migration or reversed migration.
Section \ref{sec:conclusion} summarizes the major conclusion.

\section{Models}
\label{sec:model}

In this section, we describe the disk profile adopted for three series of  simulations of terrestrial planet formation. In Section \ref{subsec:solid}, the solid disk profile declared is employed to produce initial embryos and planetesimals for entire simulations. The isothermal gaseous disk in Section \ref{subsec:iso} is related to inward migration, while the non-isothermal disk (see Section \ref{subsec:noniso}) corresponds to reversed migration.

\subsection{Solid disk profile}
\label{subsec:solid}

Generally, planets are formed in the protoplanetary disk made up of solid and gaseous materials.
In the core-accretion model, solid particles grow from sub-micron sized dust to pebbles of meter scale and then form planetesimals in size from sub-km to 100 km.
Through subsequent runaway coagulation, embryos could grow up to $10^3$ km with a mass of 0.1 $M_{\oplus}$.
In our simulations, we assume the solid disk may have already evolved over million years, which is composed of embryos and planetesimals.
The total mass of solid particles constrains the disk evolution and planet formation, and the optimal way to evaluate is the luminosity of the mm continuum emission.
However, the observation of photometry can solely give the lower bounds of disk solid mass or the mass of residual disk \citep{Greaves2010, Najita2014, Manara2018}.
An alternative widely used technique is to sum up the masses of rocky planets and cores of giant planets in planetary systems.
Table \ref{tab:multi} lists the parameters of thirty-one multiple systems made up of rocky planets around M dwarfs, where the ratios of the solid mass of disk (the summation of mass of the planets) and M dwarfs vary from 0.0035\% to 0.02\%. Thus, this may provide substantial clue to assume the initials for simulations of planetary formation.

The solid surface density ($\Sigma_s$) scales with heliocentric distance $r$ as $\Sigma_s \propto r^{-k}$, where $k$ is the power-law disk slope.
The empirical MMSN model suggests $\Sigma_s \propto r^{-3/2}$, which was ever believed to be $\Sigma_s \propto r^{-1.6}$ to produce close-in super-Earths \citep{Chiang2013}. Furthermore, \citet{Raymond2014} addressed that the solid surface density spans a wide range from $\Sigma_s \propto r^{0.5}$ to $r^{-3.2}$ to satisfy the diversity of planetary systems.
Here we adopt three typical profiles for the disk surface density $\Sigma_s \propto r^{-3/2}$, $r^{-2}$, and $r^{-5/2}$ to explore the essential role that may play a part in the accretion and evolution.

\subsection{Isothermal gaseous disk}
\label{subsec:iso}

When the planetary embryos are embedded in viscous disk, the angular momentum exchange with gaseous disk promotes density waves near the location of Lindblad and corotation resonances.
Protoplanets migrate inward or outward depending on the total torque
\begin{equation}
\begin{aligned}
&\Gamma_{\rm tot} = f_{\rm tot} \Gamma_0/ \gamma, \\
&\Gamma_0 = (q/h)^2 \Sigma_p r_p^4 \Omega_p^2,
\end{aligned}
\label{Equ_tot}
\end{equation}
where $f_{\rm tot} = f_{\rm LB}+f_{\rm CR}$, and $f_{\rm tot}$, $f_{\rm LB}$ and $f_{\rm CR}$ are the coefficient for total, Lindblad and corotation torque, respectively.
$\gamma = 1.4$ is the adiabatic index, and $q=m_p/M_{\ast}$ is the mass ratio between embryos and their host star.
$r_p$, $h=H/r_p$, and $\Omega_p=\sqrt{GM_{\ast}/r_p^3}$ indicate the distance from star, aspect ratio of disk and the Keplerian angular velocity at $r_p$, respectively.

We adopt the empirical MMSN model with the surface density of gaseous disk at distance $a$ given by \citep{Hayashi1981}
\begin{equation}
    \Sigma_g = \Sigma_0 \ (\frac{a}{1 \ \rm{AU}})^{-k} \ \exp(-\frac{t}{t_n}) \ \rm{g \ cm^{-2}},
    \label{eq:sigma}
\end{equation}
where $\Sigma_0 = 1700 \ \rm{g \ cm^{-2}}$ and $t_n = 10^6 \ \rm{yr}$ represent the initial gaseous disk density at 1 AU and the depletion timescale, respectively.
The power-law index $k$ is in line with the solid disk.
At the inner region of the gaseous disk, it would be truncated due to the stellar magnetic field near the corotation radius \citep{Koenigl1991}

\begin{equation}
\label{eq:r_m}
\begin{aligned}
    r_{\rm m} = &(7.48 \times 10^{-4}) \beta' (\frac{R_{\ast}}{0.1 R_{\odot}})^{12/7} (\frac{B_{\ast}}{180 \ G})^{4/7}  (\frac{M_{\ast}}{0.1 M_{\odot}})^{-1/7} \\
     & \times (\frac{\dot{M_{\ast}}} {10^{-10} M_{\odot} yr^{-1}})^{-2/7} \rm{AU} ,
\end{aligned}
\end{equation}
where $\beta' \le 1$ is a free parameter, $R_{\ast}$, $B_{\ast}$, $\dot{M_{\ast}}$ refer to the radii, magnetic field and accretion rate of the central star, respectively.

In an isothermal gaseous disk, the stellar irradiation is assumed to be the dominant heating source \citep{Garaud2007}.
The planetary embryos usually undergo type I inward migration until gas depleted or they are trapped into inner hole.
The timescales of migration and eccentricity damping are given by \citep{Tanaka2002, Tanaka2004, Cresswell2006}
\begin{equation}
\label{eq:migI}
\begin{aligned}
\tau_{\rm migI}=\frac{r}{|\dot{r}|} &= \frac{m_p \sqrt{GM_{\ast} r}}{2\Gamma_{\rm total}}  \\
&\simeq \frac{1}{f_1(2.7+1.1\beta)}\left(\frac{M_{\ast}}{m_p}\right)\left(\frac{M_{\ast}}{\Sigma_g r_p^2}\right)\left(\frac{H}{r_p}\right)^2\\
&\times
\left[\frac{1+(\frac{er_p}{1.3h})^5}{1-(\frac{er_p}{1.1h})^4}\right]\Omega_p^{-1}
\ \rm{yr},
\end{aligned}
\end{equation}
and
\begin{equation}
\label{equ:edamp}
\begin{aligned}
\tau_{\rm e}=\left(\frac{e}{\dot{e}}\right)_{\rm edamp} \simeq & \frac{Q_e}{0.78}\left(\frac{M_{\ast}}{m_p}\right)\left(\frac{M_{\ast}}{r_p^2\Sigma_g}\right) \\
&\times \left(\frac{h}{r_p}\right)^4\Omega_p^{-1}\left[1+\frac{1}{4}\left(e\frac{r_p}{h}\right)^3\right]
\ \rm{yr},
\end{aligned}
\end{equation}
respectively, where $f_1$ is a reduction factor of migration speed. $\beta = -\mathrm{d} \ln \Sigma_g / \mathrm{d} \ln r_p$ and $Q_e = 0.1$ is the coefficient.

\subsection{Non-isothermal gaseous disk}
\label{subsec:noniso}
While for the self-consistent non-isothermal protoplanetary disk, the thermal structure is determined by viscous heating at the inner optically thick regions and stellar irradiated for the outer optically thin region \citep{Garaud2007, Paardekooper2011}.
Following the disk temperature profile of \citet{Liu2015}, the transition radius between two different heating regimes is simplified as
\begin{equation}
    r_{\rm{trans}} \simeq 0.06 \ (\frac{M_{\ast}}{0.08 \ M_{\odot}})^{0.856} \      (\frac{\alpha_{\nu}}{10^{-3}})^{-0.36} \ (\frac{\kappa_0}{0.02})^{0.36} \ \rm{AU},
    \label{eq:r_trans}
\end{equation}
where $\alpha_{\nu} = 10^{-3}$ and $\kappa_0 \simeq 0.02$ are the disk turbulent efficiency factor and disk opacity, respectively.

The entire torque in Eqn. \ref{Equ_tot} for embryos can be expressed as \citep{Wang2021}
\begin{equation}
f_{\rm tot} = f_{\rm nsc} Q(q) + f_{\rm LB}
\end{equation}

\begin{equation}
f_{\rm nsc} = coef \times (1 - \frac{2(\frac{r_p}{r_{\rm trans}})^2}{1+(\frac{r_p}{r_{\rm trans}})^2}),
\end{equation}

\begin{equation}
Q(q) = \frac{(\frac{q}{q_1})^2 (\frac{q}{q_2})^{-2}}{2 (\frac{q}{q_1})^2 (\frac{q}{q_2})^{-2} + (\frac{q}{q_1})^2 +(\frac{q}{q_2})^{-2}},
\end{equation}

\begin{equation}
f_{\rm LB} = 1 - \frac{2 (\frac{r_p}{r_{\rm m}})^2}{1 + (\frac{r_p}{r_{\rm m}})^4},
\end{equation}
where $f_{\rm nsc}$ and $Q(q)$ are the coefficient and saturation parameter for the fully non saturated component of the corotation torque, respectively.
$coef$ is the strength factor of the corotation torque.
Here we take $coef = 5$, $q_1 = 3 \times 10^{-7} \, M_{\odot}$ and $q_2 = 10^{-3} \, M_{\odot}$ in our simulations.

The type I migration and eccentricity damping timescales of bodies in non-isothermal disk \citep{Tanaka2002, Kley2012, Liu2015, Wang2021} are
\begin{equation}
    \tau'_{\rm migI} = \frac{m_p \sqrt{GM_{\ast}r}}{2\Gamma_{\rm tot}} = \frac{m_p \gamma \sqrt{GM_{\ast}r_p}}{2 f_{\rm tot} (\frac{q}{h})^2 \Sigma_g r_p^4 \Omega_p^2} \ \rm{yr},
    \label{eq:ta2}
\end{equation}
and
\begin{equation}
    \tau'_e = h^2 f_{\rm tot} \tau'_{\rm migI} = \frac{m_p h^2 \gamma \sqrt{GM_{\ast}r_p}}{2 (\frac{q}{h})^2 \Sigma_g r_p^4 \Omega_p^2} \ \rm{yr},
    \label{eq:te2}
\end{equation}
respectively.

The acceleration of the embryos with mass $m_i$ is
\begin{equation}
\begin{aligned}
\frac{d}{dt}\mathbfit{V}_i = -\frac{G(M_*+m_i)}{{r_i}^2}\left(\frac{\mathbfit{r}_i}{r_i}\right) +\sum _{j\neq i}^N Gm_j \left[\frac{(\mathbfit{r}_j-\mathbfit{r}_i)}{|\mathbfit{r}_j-\mathbfit{r}_i|^3}- \frac{\mathbfit{r}_j}{r_j^3}\right]&\\
+\mathbfit{F}_{\rm damp}+\mathbfit{F}_{\rm migI}&,
\end{aligned}
\end{equation}
where $\mathbfit{V}_i$ and $\mathbfit{r}_i$ are the velocity vectors and positions of embryos $i$ in the stellar-centric coordinates, respectively.
The external forces are defined as
\begin{equation}
\begin{aligned}
&\mathbfit{F}_{\rm damp} = -2\frac{(\mathbfit{V}_i \cdot \mathbfit{r}_i)\mathbfit{r}_i}{r_i^2 \tau^{'}_{\rm e}} \ \rm{(or} \; \tau_{\rm e} \; \rm{for \; inward \; migration)},\\
&\mathbfit{F}_{\rm migI}=-\frac{\mathbfit{V}_i}{2\tau^{'}_{\rm migI}} \ \rm{(or} \; \tau_{\rm migI} \; \rm{for \; inward \; migration)}.
\end{aligned}
\end{equation}

\begin{table}
\caption{Multi-planetary systems around M dwarfs.}
\label{tab:multi}
 \begin{tabular}{ccccc}
 \hline
 Number & Host star & Stellar mass & $N_{\rm p}$ & $M_{\rm sys}/M_{\ast}$ \\
  & & ($M_{\odot}$) &  & ($\times 10^{-4}$) \\
  \hline
1 & GJ 1061 & 0.120 & 3 & 1.19 \\
2 & GJ 1132 & 0.180 & 2 & 0.71 \\
3 & GJ 180 & 0.440 & 3 & 1.27 \\
4 & GJ 229 A & 0.581 & 2 & 0.81 \\
5 & GJ 3323 & 0.164 & 2 & 0.79 \\
6 & GJ 3473 & 0.360 & 2 & 0.77 \\
7 & GJ 357 & 0.346 & 3 & 0.98 \\
8 & GJ 3998 & 0.517 & 2 & 0.51 \\
9 & GJ 667 C & 0.330 & 5 & 1.76 \\
10 & GJ 682 & 0.270 & 2 & 1.45 \\
11 & GJ 887 & 0.489 & 2 & 0.35 \\
12 & K2-133 & 0.460 & 4 & 0.89 \\
13 & K2-146 & 0.310 & 2 & 1.30 \\
14 & K2-18 & 0.500 & 2 & 1.00 \\
15 & K2-239 & 0.400 & 3 & 0.27 \\
16 & K2-240 & 0.580 & 2 & 0.50 \\
17 & K2-3 & 0.550 & 3 & 0.62 \\
18 & Kepler-138 & 0.540 & 6 & 0.15 \\
19 & Kepler-42 & 0.150 & 3 & 1.18 \\
20 & Kepler-445 & 0.280 & 3 & 1.94 \\
21 & Kepler-446 & 0.190 & 3 & 1.65 \\
22 & L 98-59 & 0.290 & 3 & 0.59 \\
23 & LHS 1140 & 0.180 & 2 & 1.43 \\
24 & LTT 3780 & 0.400 & 2 & 0.84 \\
25 & TOI-270 & 0.400 & 3 & 1.04 \\
26 & TOI-700 & 0.420 & 3 & 0.83 \\
27 & TOI-776 & 0.540 & 2 & 0.51 \\
28 & TRAPPIST-1 & 0.080 & 7 & 2.01 \\
29 & Teegarden's & 0.090 & 2 & 0.73 \\
30 & Wolf 1061 & 0.290 & 3 & 1.33 \\
31 & YZ Cet & 0.130 & 3 & 0.56 \\
\hline
 \end{tabular}
 \begin{tablenotes}
    \footnotesize
    \item[]Note: $N_{\rm p}$ and $M_{\rm sys}$ are the number and total mass of planets in planetary systems.
    \end{tablenotes}
 \end{table}

Here we ignore the type II migration of protoplanets because the gap-opening mass in our disk model given by $M_{\rm gap} \simeq 40 \alpha_{\nu} (\frac{h}{a})^2 M_{\ast} \simeq 10^{-4} M_{\ast}$ is comparable to the total mass of the solid disk \citep{Ida2004}.

\section{Numerical simulation}
\label{sec:simulation}

To extensively understand the accretion and growth of embryos and planetesimals with self-gravity in protoplanetary disk, we conduct numerical simulations with MERCURY6 package \citep{Chambers1999} by considering three diverse models of in-situ formation, inward migration and reversed migration.
In the following, we brief the numerical setup for the simulations of planetary formation. As shown in Figure \ref{fig:a_M}, the masses of M dwarfs and host stars range from 0.075 to 0.6 $M_{\odot}$, therefore we assume that the stellar mass is adopted to be 0.08, 0.2, 0.4 and 0.6 $M_{\odot}$, respectively, in our simulations. Moreover, as to the disk profile, the mass of solid disk is set to be 0.01\% of the mass of central star, which infers from the mass ratio between planets and host stars for each system in Table \ref{tab:multi}. And we adopt three disk slopes of $k= 3/2$, 2 and $5/2$, respectively, as described in Section \ref{subsec:solid}. The initial parameters for each run are substantially summarized in Table \ref{tab:param}.
The initial masses of embryos and planetesimals are set to be 0.275 and 0.005 $M_{\oplus}$, respectively.
The number of bodies for each simulation depends on the entire mass of solid disk, which is denoted as $N_{\rm emb}$ and $N_{\rm plt}$ (see Table \ref{tab:param}) for embryos and planetesimals, respectively.
The embryos and planetesimals are assumed to be in near-circular coplanar orbits, and the relative velocity of bodies is given by $v = \sqrt{e^2 + i^2}~v_k$, where $v_k$ is the local Keplerian velocity and $e=2i$ \citep{Ida1992}. Here we adopt the initial eccentricity $e \le 0.0005$ and inclination $i \le 0.001\degr$. The argument of pericenter, longitude of ascending node and mean anomaly are randomly generated between 0\degr and 360\degr. Furthermore, we take the disk that consists of embryos and planetesimals spreading from 0.01 to 0.5 AU, on the basis of the truncation radius in Eqn.\ref{eq:r_m} and the orbital distribution of low-mass planets shown in Figure \ref{fig:a_M}.
The time step for each integration is set to be less than 1/30 of the orbital period of the innermost planet and the accuracy is given as $10^{-15}$. The integration timescales last 1 to 10 million years for each. Thus, there are totally 66 runs that we performed, which cost the computation time of over 50 thousand CPU hours.

\subsection{In-situ formation}

\begin{figure}
\includegraphics[width=\columnwidth]{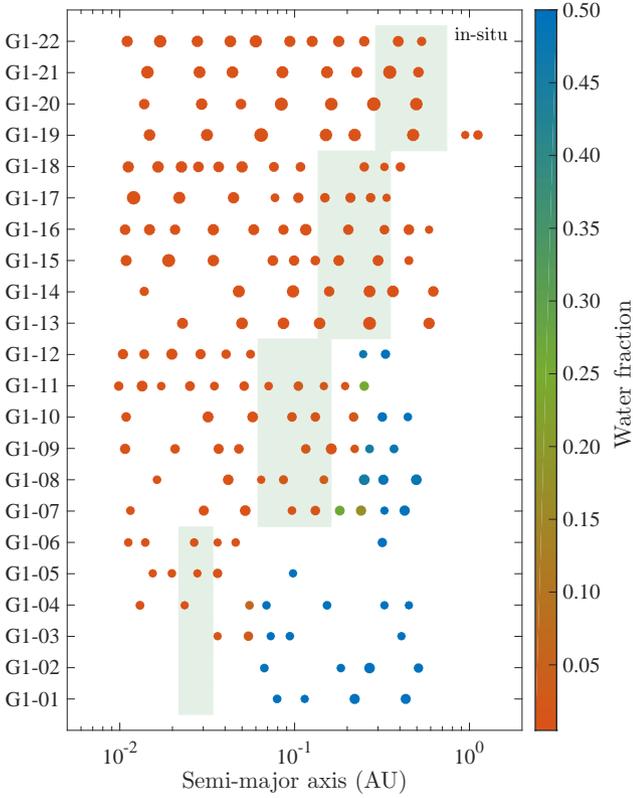}
\caption{The configuration of planetary systems formed by in-situ simulations from G1-01 to G1-22. The shaded area is the habitable zone of each case. Size and color show the mass and water fraction of planets, respectively.
\label{fig:insitu}}
\end{figure}

To investigate the influence of the disk and stellar mass on in-situ formation of terrestrial planets, we perform 22 simulations based on various solid disk profiles around the stars.
The initial parameters and outcomes are listed as G1-01 to G1-22 in Table \ref{tab:param}, and the final configuration of each planetary system is shown in Figure \ref{fig:insitu}, where the size and color of symbols represent the mass and water fraction of planets, respectively. Only planets above 0.3 $M_{\oplus}$ are presented.

The simulation results show that more planets with tightly compact structure would be formed in steeper disk as expected by \citet{Raymond2005}.
Although the maximum mass of planet increases along with the disk slope, the average mass of planets goes down, which means the mass distribution is very diverse.
\citet{Bolmont2014} proposed that the planets formed at the inner disk with steeper density profile are usually more massive.
The growth of planets is directly related to accretion efficiency, which is referred to the ratio that the planetesimals are accreted by embryos.
However, there is no significant correlation between the disk slope and accretion efficiency.
The likelihood of planets embedded in the habitable zone may grow with the power-law index of disk density.
The steeper disk profile results in the closer planets. As a result, the less water is retained in the planets owing to the extremely low water delivery through in-situ formation \citep{Raymond2005, Bolmont2014}.
On the other hand, with the growth of stellar mass, the disk mass and the number of the bodies go up so that the timescale of planetesimal accretion turns to be longer and the planetary formation becomes less efficient.
The accretion efficiency, census as well as the maximum mass of final planets can ascend when the stellar mass increases.
Benefiting to the receding of ice line with an increasing stellar mass, the planets have lower water contents and unable to be habitable about massive stars.

In the simulations, approximately 55-100\% of planetesimals are accreted and on average $7.77^{+3.23}_{-3.77}$  planets with mass of $1.23^{+4.01}_{-0.93} \ M_{\oplus}$ could be formed.
The Earth-mass planets can grow easily via in-situ accumulation, probably providing a suitable scenario to elucidate the formation of planetary system composed of a vast number of the planets.
The resultant configuration depends significantly on the initial positions where the embryos are generated.

\begin{figure}
\includegraphics[width=\columnwidth]{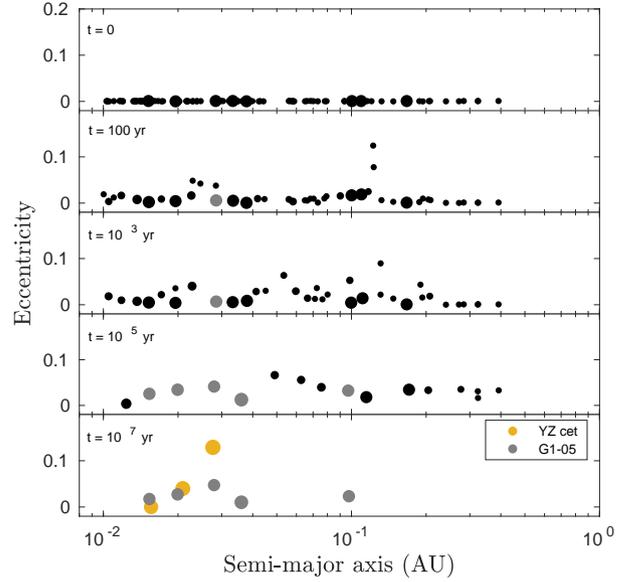}
\caption{Snapshots of an in-situ accumulation for G1-05. \emph{Four upper panels}: time evolution of the bodies (small/large black circles for  planetesimals/embryos). \emph{Bottom panel}: the formed planets (grey dots) with mass above 0.3 $M_{\oplus}$. Yellow points show the configuration of YZ cet planetary system. The symbol size scales with the planetary mass.
\label{fig:insitu_008}}
\end{figure}

Comparing the typical lifetime of the gaseous disk dissipation with the coagulation and accretion of solid bodies, we assume the embryos and planetesimals have formed in core accretion and the migration terminated because of the dissipated gaseous disk.
Here in our simulations, we simply take into consideration  mutual interaction among the embryos and planetesimals, which can accumulate each other in situ.
Figure \ref{fig:insitu_008} shows the snapshots of in-situ scenario for case G1-05, where the power-law index for the disk slope is $k=5/2$ and the stellar mass is 0.08 $M_{\odot}$.
In the very beginning, eight embryos (marked out by larger black dots) and 94 planetesimals (by small black dots) are initially embedded in solid disk, and then accumulate and accrete intensely in the first thousand years.
Subsequently, the eccentricities of planetesimals have been remarkably stirred up until the collisions settle down slowly at about 10 Myr.
Approximately 70.21\% of planetesimals are accreted by planetary embryos or accumulated to produce protoplanets over the evolution, which the accretion efficiency is given in Table \ref{tab:param}.
Finally, five terrestrial planets with mass of 0.315, 0.335, 0.31, 0.585 and 0.315 $M_{\oplus}$ are formed at 0.0153, 0.0199, 0.0279, 0.0360 and 0.0976 AU, respectively.
Each body revolves the host star in near circular and co-planar orbits with a very low eccentricity (< 0.05) and inclination ($<1.5\degr$).
The separations of adjacent planets in this simulation are larger than thirty times of mutual Hill radius, thereby leading to secular stable orbits.
Interestingly, we observe that four inner planets appear to be trapped into a chain resonance of near 3:2, 5:3 and 3:2 mean motion resonances (MMRs) \citep{Mills2016, Luger2017, Christiansen2018, Shallue2018}.
The configuration of three inner planets in the synthetic planetary system is somewhat comparable to those in the YZ cet system, in which three Earth-like planets (0.75, 0.98 and 1.14 $M_{\oplus}$) locate at 0.0156, 0.0209 and 0.0276 AU away from host star \citep{Astudillo-Defru2017}.
The final planetary masses between our simulations and the observed system differ in that a more massive solid disk or higher accretion efficiency is required.
Moreover, we estimate the HZ \citep{Kopparapu2013} of the synthetic system, being indicative of the third planet may be habitable if water is delivered to the body.

\subsection{Inward migration formation}

Rocks embedded in gaseous disk tend to orbit near Keplerian velocity \citep{Weidenschilling1977}.
The difference between the velocity of rocks and gas gives rise to the angular momentum exchange and radial drift.
When protoplanets are formed, density waves would be excited in three-dimensional isothermal gaseous disk and the bodies undergo type I migration within a very short timescale \citep{Tanaka2002}.
Here we explore the formation of rocky planets orbiting low-mass stars by accounting for mutual interaction with gaseous disk under the accretion.
We adopt a reasonable reduction factor $f_1 = 0.3$ in Eqn. \ref{eq:migI} \citep{Pan2020}, where the gas accretion of planets is not considered.

For inward migration scenario, we carry out dozens of simulations to explore the planetary formation in the combination of diverse disk density profile and stellar mass.
Table \ref{tab:param} summarizes the initials and typical results for G2-01 to G2-22.
When disk slope goes steeper, a great number of solid bodies will be trapped in the inner disk, thus the interior planetary embryos may have more opportunity to accrete and embrace the nearby planetsimals.
The averaged accretion efficiency for inward migration is roughly 88\%, which  closely resembles that of in-situ formation.
Super-Earths can form easily around stars with mass of 0.08 $M_{\odot}$ due to inward migration, with a maximum mass up to 2.53 $M_{\oplus}$  as shown in G2-01.

\begin{figure*}
\centering
\includegraphics[width=\linewidth]{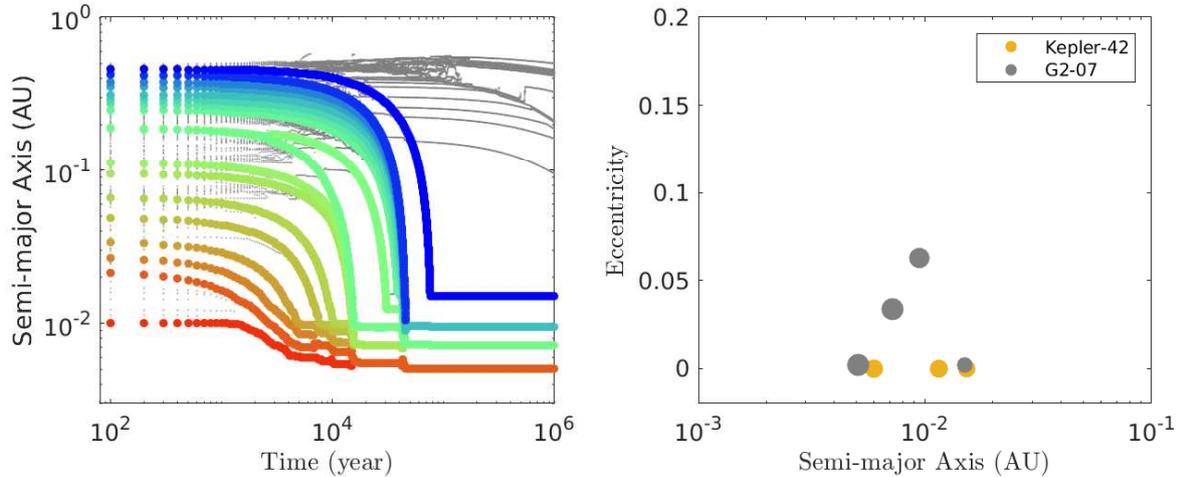}
\caption{Inward migration of planet formation for G2-07. \textit{Left panel}: the evolution under gaseous disk. The color and gray points denote accretion and evolution of embryos and planetesimals, respectively. \textit{Right panel}: the comparison of the simulation with Kepler-42 system. The symbol size is in proportion to planet radius.
\label{fig:008mig}}
\end{figure*}

\begin{figure*}
%%\centering
  \subfigure[]{\includegraphics[scale=0.4]{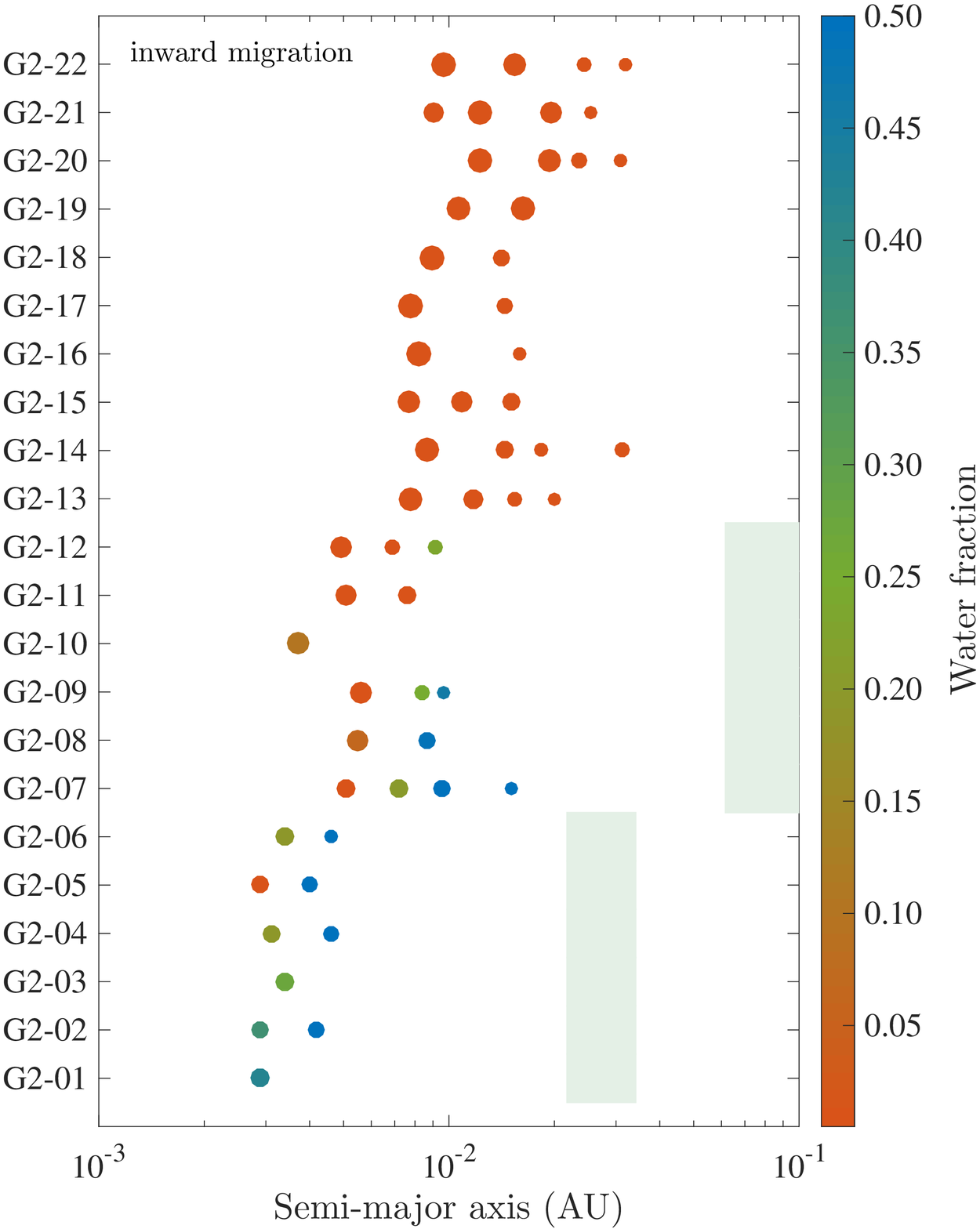}}
  \subfigure[]{\includegraphics[scale=0.4]{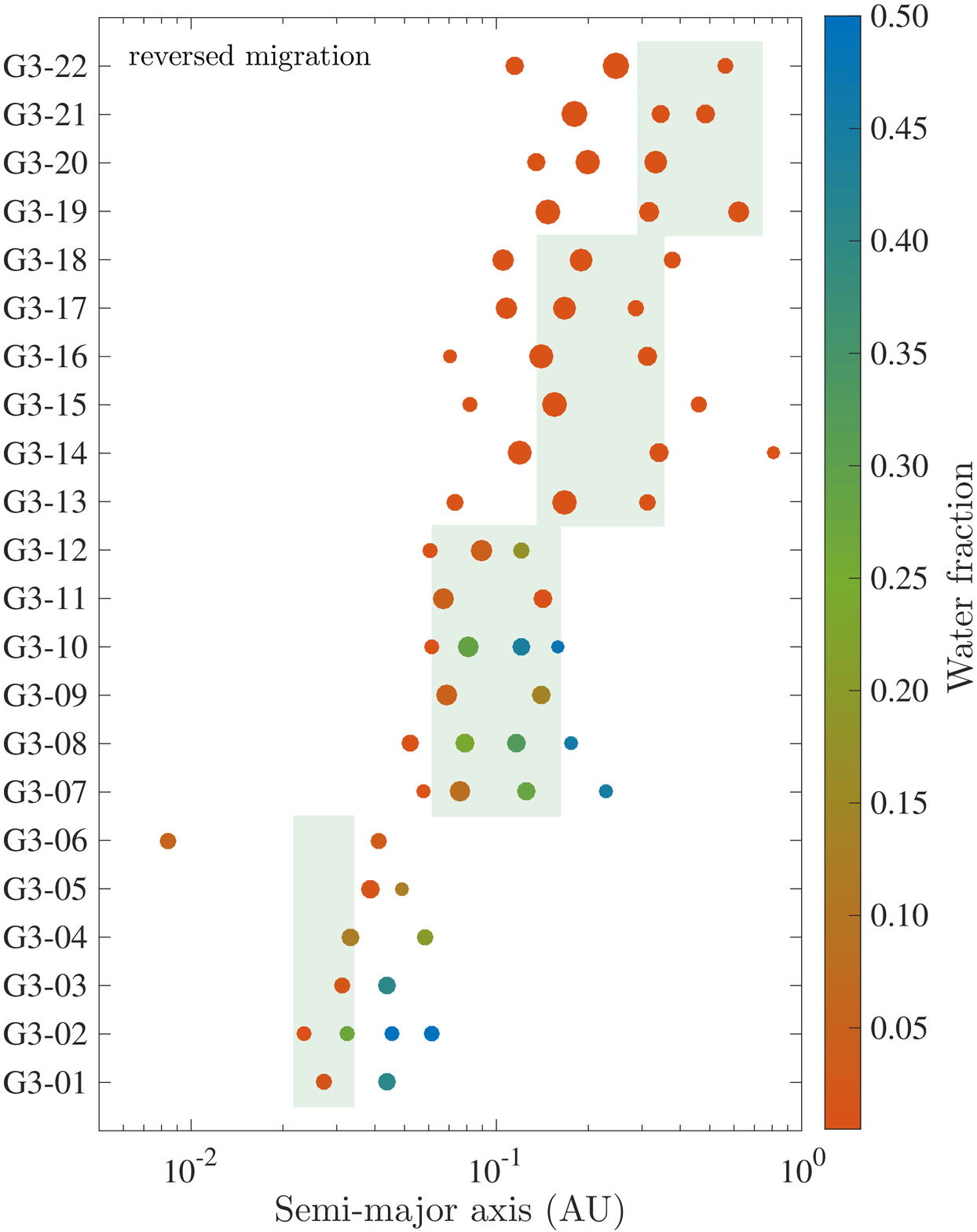}}

\caption{Same as Fig. \ref{fig:insitu}. (a) Inward migration: simulations for G2-01 to G2-22. (b) Reversed migration: simulations for G3-01 to G3-22. }
\label{fig:inward}
\end{figure*}

As the materials in solid disk rises with the host star, the embryos can accrete more planetesimals and accumulate each other intensely into larger planets.
When the disk profile becomes steeper, corresponding to the stellar masses of 0.4 or 0.6 $M_{\odot}$, hot sub-Neptunes are observed to have formed in our simulations. The sub-Neptunes might accrete into gas-giants after runaway growth of the envelope, providing a likely explanation for higher occurrence rate of hot Jupiter around massive stars.
Our simulations present that roughly $2.55^{+1.45}_{-1.55}$ terrestrial planets with an averaged mass of $3.76^{+8.77}_{-3.46}~M_{\oplus}$ eventually survive and orbit M dwarfs after migration and gas dissipation, and they are trapped into chain resonances like near 8:5, 3:2 and 2:1 MMRs (Figure \ref{fig:inward} (a)).
The interaction with gaseous disk in classical migration scenario drives the bodies move inward until disk disperses.
While the migration timescale is theoretically far less than that of  dissipation in the gaseous disk, thus in our simulations the planets halt migrating near the inner boundary of disk spreading from 0.0029 to 0.0311 AU.
Here we come to conclusion that only close-in terrestrial planets and hot Neptunes might be well explained by inward migration model.

Figure \ref{fig:008mig} shows a typical run of  terrestrial planet formation for G2-07 via inward migration, where the stellar mass is 0.2 $M_{\odot}$ and the power-law index for disk profile is $k=3/2$.
The color and gray points represent the time evolution of semi-major axis for the embryos and planetesimals, respectively.
The planetesimal accretion occurs mainly before the termination of migration close to 50,000 yr, whereas the embryos (marked out by color symbols) accumulate around the inner edge of disk as shown in left panel of Figure \ref{fig:008mig}.
Moreover, for G2-07, we observe that four terrestrial planets with mass of 2.345, 2.320, 1.490 and 0.300 $M_{\oplus}$ are formed after migration at $a=$ 0.0051, 0.0072, 0.0095 and 0.015 AU, respectively. It is noteworthy to mention that the resultant planets are trapped in 5:3, 3:2 and 2:1 resonance chain when the gaseous disk dissipated.
Furthermore, we conduct additional simulations for this system, which reveal that four terrestrial planets can sustain stable over the timescale up to $10^8$ yr.
The right panel in Figure \ref{fig:008mig} exhibits the comparison of this synthetic planetary system of G2-07 and the Kepler-42 system (with the stellar mass of 0.144 $M_{\odot}$), which harbors three planets at the orbit of 0.006, 0.0116 and 0.0154 AU \citep{Muirhead2012}, respectively.
The radii of planets are a little smaller than those of the simulation with 0.78, 0.73 and 0.57 times of Earth radius \citep{Morton2016}.
The deviation of planetary radii between the simulation and the Kepler-42 system might be induced by their variational solid disk masses in relation to the mass of host stars.
On the other hand, the final configuration of the simulation produced a similar distribution of eccentricities with Kepler-42 system, in which planets are assumed in circular orbits due to transit observation.

\subsection{Reversed migration formation}

To in-depth understand the aggregation of planets near the inner edge of disk and clarify the observed distribution of planets around M dwarfs, here we apply non-isothermal gaseous disk during planetary migration.
From interior viscously heated to exterior radiation heating, embryos tend to assemble where inward and outward migration torque are balanced, which is also near the transition radius of two heating mechanisms \citep{Garaud2007}. The results of case G3-01 to G3-22 are shown in  Figure \ref{fig:inward} (b)  (see also Table \ref{tab:param}).

\begin{figure*}
\centering
 \includegraphics[width=\linewidth]{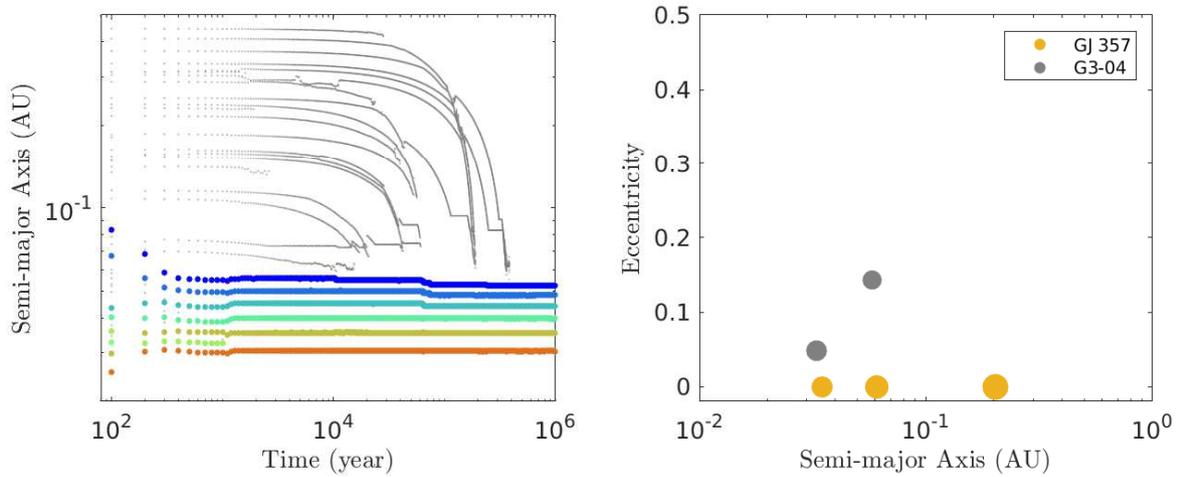}
 \caption{Similar to Fig. \ref{fig:008mig}. Reversed migration for G3-04. The final configuration is compared with the GJ 357 system.
 \label{fig:G3-4}}
 \end{figure*}

In the reversed migration model, the simulation outcomes suggest that the number of formed planets in final systems usually ranges from 2 to 4, averagely $2.85^{+1.15}_{-0.85}$.
Neither single-planet system nor multiple systems with planets more than five is found in this scenario, which is different from those of \citet{Alibert2017} that showed 76\% systems own only one planet.
Systems with planets up to 15 could be formed in a broader scale of disk $\sim$ 10-100 AU \citep{Miguel2020}.
When the stellar mass goes up to 0.6 $M_{\odot}$, the planetary embryos can even accrete more bodies and become a Neptune-like planet.
In such circumstance, the planets no longer pile up and get that close to their host stars but range from 0.0084 to 0.56 AU.
The eccentricity of planets might be excited up to 0.4 through gravitational encounters, while the configuration of planetary systems may become looser with a maximum value of period ratio up to 10.
The results in relation to the inner region of non-isothermal gaseous disk may throw light on the observed distribution of planets orbiting M dwarfs and provide a possible mechanism to explain formation of moons.

Figure \ref{fig:G3-4} shows the typical evolution of reversed migration formation for case G3-04. From  Eqn. \ref{eq:ta2}, the migration timescale of bodies is evaluated to be inversely proportional to the mass, such that planetesimals complete drifting at about two or more orders of magnitude later than embryos as shown in Fig. \ref{fig:G3-4}.
When they stop migrating, six planets are formed in the compact system.
Intense impact occurred over the dynamical evolution when gaseous disk dispersed, thus two planets survive at approximately 0.0333 and 0.0584 AU away from the star.
The final geometry of G3-04 resembles the GJ 357 system except for the absence of planet d \citep{Luque2019}.
Here we give a likely interpretation that super-Earths might have formed in the outer disk and then undergo migration into the inner disk.

\begin{figure}
\centering
\includegraphics[width=\columnwidth]{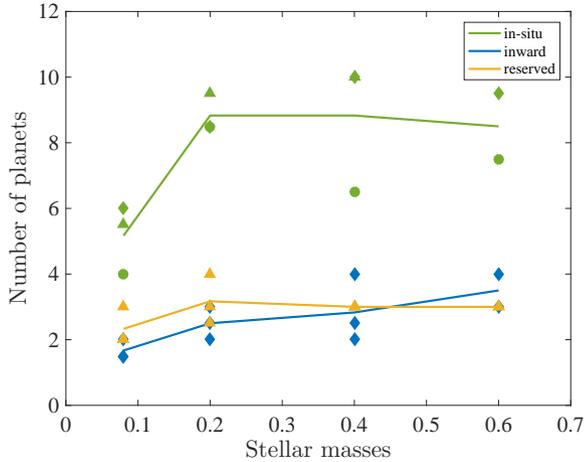}
\caption{The number of planets formed in simulations around different stellar masses. The green, blue and yellow lines show the average results of in-situ, inward migration and reversed migration models, respectively. The circles, diamonds and triangles present different disk slope with $k=3/2$, 2, $5/2$, respectively.
\label{fig:num}}
\end{figure}

\begin{figure}
\centering
\includegraphics[width=\columnwidth]{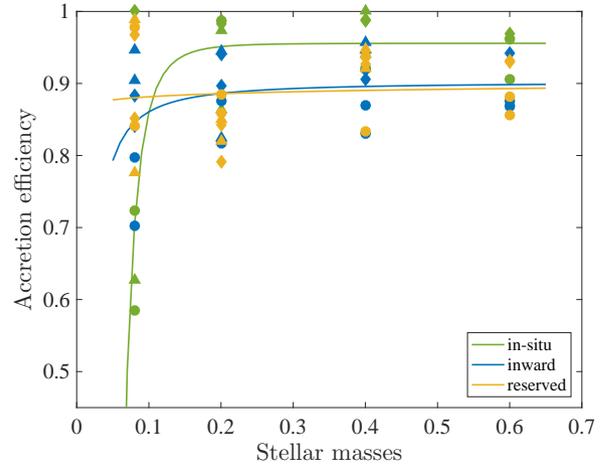}
\caption{Similar to Figure \ref{fig:num} but for the accretion efficiency in each simulation. The green, blue and yellow lines are the fitted curve of the relation between stellar mass and accretion efficiency for in-situ, inward migration and reversed migration models, respectively.
\label{fig:acc}}
\end{figure}

By comparing our simulation results with the above-mentioned models, the in-situ scenario can produce the number of planets up to 11, as the green symbols shown in Figure \ref{fig:num}.
If there is no interaction with gaseous disk, the embryos move so sightly that only planetesimals in unstable region (within 3.5 times of Hill radius) can be accreted.
Blue and yellow symbols denote the number of formed planets in the simulations under inward and reversed migration, respectively, both of which have produced approximately 2.5 planets with minimal disparity no matter how abrupt the protoplanetary disk is.
Furthermore, we show the difference in accretion efficiency of three models in Figure \ref{fig:acc}. The green, blue and yellow solid lines, respectively, represent the relation of accretion efficiency for three models versus stellar masses, each of which appears to follow a power-law approximation.
For stellar masses over 0.2 $M_{\odot}$, more than 80\% of planetesimals will be accreted by embryos.
The formation model and disk profile have very little influence on the accretion efficiency of planets except for those simulations with stellar masses of 0.08 $M_{\odot}$, where the accretion efficiency of in-situ scenario is extremely lower than that of the other two models.
We ascribe this phenomenon to the fewer solids around low-mass stars that the initial larger separation between the embryos may avoid violent collisions. Moreover, under the influence of gaseous disk, more supplies could be transported into the feeding zone of embryos thereby resulting in efficient accretion in migration scenarios.
Planets could rarely increase to 1 $M_{\oplus}$ through in-situ scenario unless the masses of host stars are larger than 0.2 $M_{\odot}$.
As to orbital distribution, the simulations reveal that ultra-short-period planets in coplanar circular orbits are commonly observed under inward migration due to eccentricity damping triggered by gaseous disk. As a comparison, the formation scenarios of in-situ and reversed migration may suffer vast collisions at the late stage of the formation, such that the planets with large eccentricity or high inclination are produced.

\subsection{Comparison between simulations and observations}

\begin{figure*}
\centering
\includegraphics[width=\linewidth]{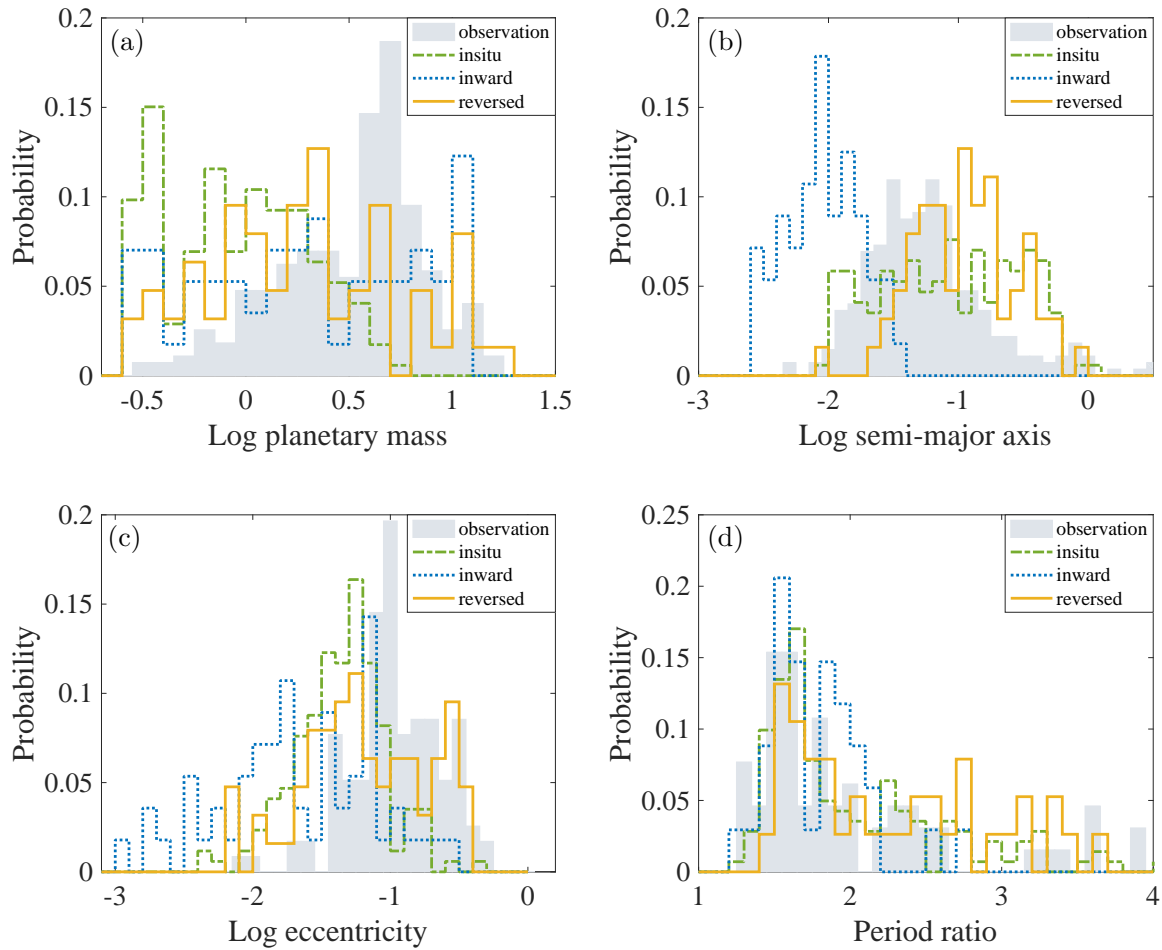}
  \caption{Comparison of the simulations and observations for planets. The distribution of mass (Panel a), semi-major axis (Panel b), eccentricity (Panel c) and period ratio (Panel d). The shadows, green dash-dotted, blue doted and yellow solid lines represent the observation, in-situ scenario, inward migration and reversed migration, respectively.
  \label{fig:compare}}
\end{figure*}

So far, hundreds of planets orbiting M dwarfs have been observed. To
clarify the characteristic of planets, we plot the distribution of
planetary mass, semi-major axis, eccentricity and period ratio of
adjacent planets with observational data that have not been corrected for bias in Figure \ref{fig:compare}, where the simulation
results from in-situ, inward migration and reversed migration are
shown as green dash-dotted, blue-dotted and yellow solid lines,
respectively, while gray shadows represent the observation
results.

The statistical results of mass distribution of observations
as shown in Panel (a) of Figure \ref{fig:compare} display that there might be a peak near 5.5 $M_{\oplus}$ and a downward trend on both sides.
Previous studies mentioned that this may be induced by the start of gas accretion that when it reaches a critical mass, the protoplanet will accrete and retain a significant gaseous atmosphere or envelope \citep{Hasegawa2014,
Jin2018}. Approximately 30\% of observed planets have 4-6
$M_{\oplus}$. The masses of planets from in-situ model are
mostly centered at the region less than 3.5 $M_{\oplus}$, while
the number of planets goes down with an increase of the planetary mass. Therefore, low-mass planets are observed in the in-situ model. In addition, the mass distribution of inward migration is comparatively
uniform with the exception of small peak at 12.5 $M_{\oplus}$.
The occurrence rate of planets
in reversed migration rises with the growth of mass and reaches the
maximum at about 2.0 $M_{\oplus}$, which coincides with the
Kepler data \citep{Gaidos2016}. The inward and reversed migration
models have more likelihood to produce massive planets because of the
higher accretion efficiency. The overall distribution trend from the
reversed model (labeled by yellow solid line) offers a good agreement with the observational distribution except the position of peaks. In addition, over 80\% of planets formed in our simulations are smaller than 3 $M_{\oplus}$, which are distinguished from the observations.
Such difference is mainly caused by two aspects: the absence of gas
accretion in our models and the initial mass of solid disk we assumed. Comparing the peaks of the observations with those from the reversed model, we suggest that solid disk with a mass more than 0.02 percentage of stellar mass may match the observations.

Without regard to the selection effect and the bias of observations, we notice nearly 89\% of samples locate in the range of [0.01 0.3] AU, and approximately 57\% accumulates between 0.03 to 0.1 AU, as seen from Panel (b) of Figure \ref{fig:compare} for the distribution of semi-major axis.
Two peaks near 0.03 and 0.06 AU are also illustrated.
No discerned features are shown in the semi-major axis distribution for in-situ
model, except nearly 40\% of planets locate in the 0.01-0.06 AU
away from host stars. For inward migration scenario, the planets terminate
migration near the inner boundary of gaseous disk with semi-major
axis below 0.04 AU. There is a peak near 0.01 AU, which may give the explanation of the formation of hot
planets. As a comparison, the reversed migration scenario produces a  resembling distribution in semi-major axis with the observations, where $\sim$
79\% of planets reside between 0.02 to 0.3 AU. However, the position of
peak is slightly larger than that of the observations, implying the heating
transition region in disk might be much closer than the temperature
structure we assume.

Planets in proximity to stars are commonly observed in or near circular
orbits. The eccentricity distribution of observations in
Panel (c) of Figure \ref{fig:compare} shows a peak near $e \sim 0.1$.
The percentage of planets with an eccentricity larger than 0.1 is
approximately 12.8\%, which may result from frequent
collisions and dramatic impacts in the formation. Most of planets hold
eccentricities less than 0.1 due to the low
accretion efficiency in the in-situ model. With the interaction of gaseous
disk, the eccentricities are gradually damped (blue dotted line in Panel (c) of Figure \ref{fig:compare}). However, a handful
of embryos gather together near the transition region through
reversed migration. Planets formed through
this model have a higher chance to be heavily stirred up,
thereby leading to the population of the planets with an eccentricity above 0.1.
Therefore, the overall trend of the distribution of eccentricity
derived from reversed migration agrees well with that of
observations.

In multiple systems, the planets tend to accumulate near MMRs
like 3:2, 5:3 and 2:1 as Panel (d) shown in Figure
\ref{fig:compare}. Comparing with the observations, the period ratio
of planet pairs formed by in-situ simulations prefer to accumulate
near 1.6 approximate to 3:2 MMR.  Nearly 85\% of planets
produced by inward migration are in tightly-packed MMR with peaks
near 3:2, 5:3 and 2:1 MMRs which is in good consistence with the
observation results.  Benefitting to the collision and mergence of
embryos and planetesimals after migration near the transition
region, the configuration of planetary systems produced by reversed
migration scenario are looser than that produced by inward
migration. Half planets locate in orbits with the period ratio more
than two. The simulation results of reversed migration are
in good agreement with the observations with an accumulation near
3:2 MMR.

To sum up, the mass of solid disk is required to be at least 0.02
percentage of stellar mass to achieve the observation distribution
of planetary mass. For those systems hosting terrestrial planets more
than four, in-situ mechanism might act as a possible scenario to explain
their formation. Inward migration model can show similar
results on the distribution of eccentricity and period ratio. The
reversed migration scenario provides a best matching with
the observations in mass, semi-major axis, eccentricity and period ratio
distributions.

\subsection{Water delivery and habitability}

The major routine for planets to harvest water is the delivery of ice by collision and mergence of embryos or planetesimals from outer region of the ice line, which is roughly estimated at $a_{\rm ice} \simeq 2.7 (L_{\ast}/L_{\odot})^{1/2}$ AU \citep{Morbidelli2000, Ogihara2009}.
The former investigation suggested that planets formed in habitable zone around M dwarfs prefer to be dry because of the inefficient water delivery in low-mass disks \citep{Lissauer2007, Raymond2007}.
Subsequently, \citet{Schoonenberg2019} showed that planets formed by pebble-driven formation scenario \citep{Ormel2017} may hold the fraction of water up to 10\%.

To explore the water delivery over the evolution of terrestrial planets, we assume the water fraction of rocks is 0.5\% by mass initially when they are interior to the ice line, whereas that fraction is 50\% for the bodies exterior to the ice line \citep{Schoonenberg2019}. The solid disks around the M Dwarfs of 0.4 and 0.6 $M_{\odot}$ are inside the ice line, so that no water delivery could happen and planets are all set to be dry in those cases.
The water loss via severe giant impact and turbulence in disk is not taken into consideration \citep{Burger2018, Liu2019}.
The fraction of water delivered from the outside to inside of ice line for each case is listed as water delivery in Table \ref{tab:param}.
The efficiency of water delivery of in-situ formation is particularly low except for case G1-07, while over 90\% of water could be transferred onto planets under migration scenarios, implying the gaseous disk may play a crucial role in the water delivery \citep{Ogihara2009}. In G1-07, the efficiency of water delivery is remarkably larger than other cases, which results from planet-planet scattering near the ice line. The resultant portion of water delivery is relatively low for G2-11 and G3-02 when compared to other simulations in migration scenario, which is caused by the dry embryos inside the ice line at the beginning and lack of the collision with the water-rich planetary embryos.
After in-situ accumulation, nearly 19.46\% of planets have a water content above 10 per cent, most of which locate outside of the ice line.
The formation of planets in HZ are quite common in gas-free circumstance, even all of them are too dry to be habitable as shown in Figure 2 \citep{Raymond2007}.
Under the delivery of water by gaseous disk, approximately one third of planets are water-rich under migration, whereas half of them are in the HZ for reversed migration scenario. However, for inward migration, the migration of planets is unsuppressed thereby triggering planets escape from the HZ.

Here we present that a significant fraction of planets around low-mass stars harbor more than 10\% water in mass \citep{Alibert2017}.
While the number of short-period Earth-like planets with more than 40\% water content is much smaller than that of \citet{Miguel2020}, in which the outer radius of disk up to 100 au is considered thereby leading to more water-rich rocks that could be transferred.
The recent observation of LHS 1140 b suggested that the water content is compatible to a maximum fraction of 10-12\% in mass, which may provide a strong support to our study \citep{Lillo-Box2020}.
In this work, the evolution of the star luminosity is not considered, in which the habitable zone may migrate inward leading to the water loss of planets \citep{Tian2015}.

\section{Conclusions and Discussions}
\label{sec:conclusion}

In this work, we investigate the formation of terrestrial planets orbiting M dwarfs by planetesimal accretion through the scenarios of in-situ scenario, inward migration and reversed migration.
We extensively conduct N-body simulations of planetesimal accretion using three models in the combination of the disk profile and stellar mass to understand the observed distribution of the planetary mass, semi-major axis, eccentricity and period ratio.

For disks with various power-law index of the density profile, we suggest more planets with smaller average mass may be produced in a steeper disk via in-situ formation \citep{Bolmont2014}, while the result is opposite in inward and reserved migration models. In addition,
the planets formed through reversed migration scenario are farther away from stars when the disk slope increases, which differs from in-situ model and inward migration. Furthermore, the steeper disk may shorten the timescale of planetary formation, but result in the water-poor planets.

The assumption of the solid disk mass growing with stellar mass brings about the formation of larger planets around massive stars.
In in-situ scenario, the number of terrestrial planets formed around massive stars is significantly larger than that of low-mass stars, while the formation of planets becomes faster around low-mass stars as \citet{Liu2020a} reported.
With the expansion of inner boundary of disk induced by strong stellar magnetic field activity, the planets seem to be much drier and prefer to terminating migration far away from the host stars as the stellar masses increase \citep{Liu2019}.
The Earth-like planets in the multiple systems could be formed easily even around the M dwarf with 0.08 $M_{\odot}$ in our simulations, especially through in-situ formation that provides a likely scenario to shed light on the formation of planetary system like TRAPPIST-1 \citep{Miguel2020}.

Comparing three kinds of mechanisms for planetary formation, in-situ accumulation generally produces $7.77^{+3.23}_{-3.77}$ terrestrial planets with an average mass of $1.23^{+4.01}_{-0.93} \ M_{\oplus}$ around M dwarfs, comparable to the outcomes of \citet{Ogihara2009} with an enhanced condensation factor.
Moreover, there are $2.55^{+1.45}_{-1.55}$ planets with mass of $3.76^{+8.77}_{-3.46} \ M_{\oplus}$ formed in the systems via inward migration, while $2.85^{+1.15}_{-0.85}$ planets with $3.01^{+13.77}_{-2.71} \ M_{\oplus}$ are yielded under reversed migration.
The planetary formation rate in reversed scenario locates between that of \citet{Alibert2017} and the outcome of \citet{Miguel2020}, who focused on the planet formation within 0.1 AU and 10 AU, respectively.
As a comparison, the results in migration scenarios are in good agreement with those estimated to be 2.5 $\pm$ 0.2 planets around each M dwarfs using Kepler observations \citep{Dressing2015}.
The reversed migration model also shows a peak near 2 $M_{\oplus}$, which is slightly larger than the results of \citet{Alibert2017} that planet radius is peaked at roughly 1 $R_{\oplus}$.
Even though, we both offer the possibility for the terrestrial planetary formation around low-mass stars with stellar mass $< 0.15 \ M_{\odot}$, which is proposed as an essential requirement by \citet{Miguel2020} for the formation of planets above Earth-mass.
A similar trend in the mass distribution with the observations and a shift of the peak in reversed migration demonstrate that nearly 80\% of planets produced in this work are less than 3 $M_{\oplus}$.
This may imply that planets larger than 3 $M_{\oplus}$ have difficulty in the formation around M dwarfs unless the mass of solid disk is twice of what we set initially in our simulation \citep{Coleman2016}.
Alternatively, those massive planets might elucidate the occurrence rate of giant planets that they may have accreted a vast fraction of gas before the gaseous disk vanished.
The unsuppressed migration in the isothermal disk and the accumulation of planets at the inner boundary of gaseous disk provide a possible way to reveal the formation of hot-planets.
The resultant semi-major axis distribution of planets in non-isothermal disk can well reproduce the observation of planets, indicating the reversed migration scenario may act as an effective mechanism to throw light on the planetary formation around M dwarfs.
Terrestrial planets formed in habitable zone are commonly observed from in-situ and reversed simulations, whereas the low efficiency of water delivery in-situ accretion leads to the planets too dry to be habitable.
If the reversed migration model is the most likely scenario to understand the formation of planets around M dwarfs, we can predict that approximately one third of planets formed under reversed migration contain over 10\% water of their masses, and half of the population locates in habitable zone.
However, the habitable planets are very limited now, and we expect that future observations can help unveil a better census of planet formation around M dwarfs.

This work demonstrates the planetesimal accretion of terrestrial planet formation around M dwarfs in protoplanetary disk.
We conclude that reversed migration may better match the observations, while the formation of planets massive than 3 $M_{\oplus}$ around low-mass stars is not yet clear \citep{Zawadzki2021}. However, recent observations show the occurrence of giant planets around M dwarfs \citep{Mercer2020}, which will lead to forthcoming investigation of the terrestrial planet formation in the presence of gas-giants and the final architecture of planetary systems around M dwarfs.

\begin{table*}
\caption{Initial parameters and typical outcomes in simulations of G1 to G3.}
\label{tab:param}
 \begin{tabular}{lccccccccccc}
 \hline
 Case No.& $M_{\ast}$ & Disk slope & Migration & $r_{\rm trans}$ & $r_{\rm ice}$ & $N_{\rm emb}$ & $N_{\rm plt}$& $N_{\rm p}$ &$M_{\rm pmax}$ & Accretion efficiency & Water delivery \\
&($M_{\odot}$)& & & (AU) & (AU) & & & & ($M_{\oplus}$) & & \\
\hline
G1-01& 0.08 & 3/2 & No & - & 0.06 & 8 & 94 & 4 & 0.960 & 72.34\% & 14.01\%\\
G1-02& 0.08 & 3/2 & No & - & 0.06 & 8 & 94 & 4 & 1.190 & 58.51\% & 0.00\% \\
G1-03& 0.08 & 2 & No & - & 0.06 & 8 & 94 & 5 & 0.630 & 57.45\% & 2.97\%\\
G1-04& 0.08 & 2 & No & - & 0.06 & 8 & 94 & 7 & 0.410 & 100.00\% & 2.97\%\\
G1-05& 0.08 & 5/2 & No & - & 0.06 & 8 & 94 & 5 & 0.585 & 70.21\% & 1.54\%\\
G1-06& 0.08 & 5/2 & No & - & 0.06 & 8 & 94 & 6 & 0.555 & 62.77\% & 0.52\% \\
G1-07& 0.2 & 3/2 & No & - & 0.24 & 20 & 234 & 9 & 1.285 & 98.72\% & 37.05\%\\
G1-08& 0.2 & 3/2 & No & - & 0.24 & 20 & 234 & 8 & 1.325 & 91.03\% & 1.31\%\\
G1-09& 0.2 & 2 & No & - & 0.24 & 20 & 234 & 9 & 1.480 & 98.29\% & 8.23\%\\
G1-10& 0.2 & 2 & No & - & 0.24 & 20 & 234 & 8 & 1.525 & 82.05\% & 5.26\% \\
G1-11& 0.2 & 5/2 & No & - & 0.24 & 20 & 234 & 11 & 1.375 & 96.58\% & 5.46\% \\
G1-12& 0.2 & 5/2 & No & - & 0.24 & 20 & 234 & 8 & 1.600 & 86.32\% & 0.57\% \\
G1-13& 0.4 & 3/2 & No & - & 0.68 & 40 & 468 & 6 & 3.550 & 92.31\% & - \\
G1-14& 0.4 & 3/2 & No & - & 0.68 & 40 & 468 & 7 & 3.000 & 92.88\% & -\\
G1-15& 0.4 & 2 & No & - & 0.68 & 40 & 468 & 9 & 3.745 & 98.93\% & -\\
G1-16& 0.4 & 2 & No & - & 0.68 & 40 & 468 & 11 & 2.040 & 98.72\% & - \\
G1-17& 0.4 & 5/2 & No & - & 0.68 & 40 & 468 & 9 & 4.720 & 100.00\% & -\\
G1-18& 0.4 & 5/2 & No & - & 0.68 & 40 & 468 & 11 & 1.980 & 94.02\% & -\\
G1-19& 0.6 & 3/2 & No & - & 1.255 & 60 & 702 & 8 & 5.240 & 96.15\% & -\\
G1-20& 0.6 & 3/2 & No & - & 1.255 & 60 & 702 & 7 & 4.840 & 90.60\% & -\\
G1-21& 0.6 & 2 & No & - & 1.255 & 60 & 702 & 8 & 4.165 & 96.87\% & -\\
G1-22& 0.6 & 2 & No & - & 1.255 & 60 & 702 & 11 & 3.140 & 96.30\%  & -\\
 \hline
G2-01& 0.08 & 3/2 & Yes & - & 0.06 & 8 & 94 & 1 & 2.530 & 70.21\% & 96.67\%\\
G2-02& 0.08 & 3/2 & Yes & - & 0.06 & 8 & 94 & 2 & 1.425 & 79.79\% & 97.80\%\\
G2-03& 0.08 & 2 & Yes & - & 0.06 & 8 & 94 & 1 & 2.300 & 84.04\% & 98.46\%\\
G2-04& 0.08 & 2 & Yes & - & 0.06 & 8 & 94 & 2 & 1.705 & 88.30\% & 97.53\%\\
G2-05& 0.08 & 5/2 & Yes & - & 0.06 & 8 & 94 & 2 & 1.665 & 94.68\% & 97.98\%\\
G2-06& 0.08 & 5/2 & Yes & - & 0.06 & 8 & 94 & 2 & 2.285 & 90.43\% & 99.20\%\\
G2-07& 0.2 & 3/2 & Yes & - & 0.24 & 20 & 234 & 4 & 2.345 & 81.62\% & 96.03\%\\
G2-08& 0.2 & 3/2 & Yes & - & 0.24 & 20 & 234 & 2 & 4.860 & 87.61\% & 95.74\%\\
G2-09& 0.2 & 2 & Yes & - & 0.24 & 20 & 234 & 3 & 5.555 & 89.74\% & 88.36\%\\
G2-10& 0.2 & 2 & Yes & - & 0.24 & 20 & 234 & 1 & 6.315 & 94.02\% & 96.10\% \\
G2-11& 0.2 & 5/2 & Yes & - & 0.24 & 20 & 234 & 2 & 4.440 & 82.48\% & 52.94\%\\
G2-12& 0.2 & 5/2 & Yes & - & 0.24 & 20 & 234 & 3 & 5.240 & 94.44\% & 90.48\%\\
G2-13& 0.4 & 3/2 & Yes & - & 0.68 & 40 & 468 & 4 & 8.680 & 86.97\% & - \\
G2-14& 0.4 & 3/2 & Yes & - & 0.68 & 40 & 468 & 4 & 9.990 & 83.12\% & - \\
G2-15& 0.4 & 2 & Yes & - & 0.68 & 40 & 468 & 3 & 6.615 & 90.60\% & - \\
G2-16& 0.4 & 2 & Yes & - & 0.68 & 40 & 468 & 2 & 12.530 & 91.88\% & - \\
G2-17& 0.4 & 5/2 & Yes & - & 0.68 & 40 & 468 & 2 & 11.920 & 95.73\% & - \\
G2-18& 0.4 & 5/2 & Yes & - & 0.68 & 40 & 468 & 2 & 11.895 & 94.66\% & - \\
G2-19& 0.6 & 3/2 & Yes & - & 1.255 & 60 & 702 & 2 & 10.330 & 87.61\% & - \\
G2-20& 0.6 & 3/2 & Yes & - & 1.255 & 60 & 702 & 4 & 10.805 & 86.75\% & -  \\
G2-21& 0.6 & 2 & Yes & - & 1.255 & 60 & 702 & 4 & 10.440 & 94.16\% & - \\
G2-22& 0.6 & 2 & Yes & - & 1.255 & 60 & 702 & 4 & 11.465 & 87.04\% & - \\
\hline
G3-01& 0.08 & 3/2 &Yes & 0.060 & 0.06 & 8 & 94 & 2 &1.590 &97.87\%& 89.38\%\\
G3-02& 0.08 & 3/2 & Yes & 0.060 & 0.06 & 8 & 94 & 4 & 0.795 & 81.91\% & 54.45\%\\
G3-03& 0.08 & 2 & Yes & 0.060 & 0.06 & 8 & 94 & 2 & 1.690 & 85.11\% & 99.29\%\\
G3-04& 0.08 & 2 & Yes & 0.060 & 0.06 & 8 & 94 & 2 & 1.635 & 96.81\% & 100.00\%\\
G3-05& 0.08 & 5/2 & Yes & 0.060 & 0.06 & 8 & 94 & 2 & 2.195 & 77.66\% & 100.00\%\\
G3-06& 0.08 & 5/2 & Yes & 0.060 & 0.06 & 8 & 94 & 2& 1.045 & 98.94\% & 100.00\%\\
G3-07& 0.2 & 3/2 & Yes & 0.130 & 0.24 & 20 & 234 & 4 & 3.660 & 88.46\% & 99.01\%\\
G3-08& 0.2 & 3/2 & Yes & 0.130 & 0.24 & 20 & 234 & 4 & 2.470 & 85.90\% & 96.95\%\\
G3-09& 0.2 & 2 & Yes & 0.130 & 0.24 & 20 & 234 & 2 & 4.150 & 79.06\% & 92.14\%\\
G3-10& 0.2 & 2 & Yes & 0.130 & 0.24 & 20 & 234 & 4 & 3.965 & 84.19\% & 98.52\%\\
G3-11& 0.2 & 5/2 & Yes & 0.130 & 0.24 & 20 & 234 & 2 & 4.190 & 85.04\% & 96.10\%\\
G3-12& 0.2 & 5/2 & Yes & 0.130 & 0.24 & 20 & 234 & 3 & 4.810 & 82.05\% & 97.14\%\\
G3-13& 0.4 & 3/2 & Yes & 0.238 & 0.68 & 40 & 468 & 3 & 10.730 & 93.80\%& - \\
G3-14& 0.4 & 3/2 & Yes & 0.238 & 0.68 & 40 & 468 & 3 & 9.415 & 83.33\%& -  \\
G3-15& 0.4 & 2 & Yes & 0.238 & 0.68 & 40 & 468 & 3 & 11.560 & 95.51\% & - \\
G3-16& 0.4 & 2 & Yes & 0.238 & 0.68 & 40 & 468 & 3 & 10.010 & 92.74\% & - \\
G3-17& 0.4 & 5/2 & Yes & 0.238 & 0.68 & 40 & 468 & 3 & 7.320 & 94.87\%& -  \\
G3-18& 0.4 & 5/2 & Yes & 0.238 & 0.68 & 40 & 468 & 3 & 6.945 & 92.09\% & - \\
%G3-19& 0.6 & 3/2 & Yes & 0.337 & 1.255 & 60 & 702 & - & - & -  \\
%G3-20& 0.6 & 3/2 & Yes & 0.337 & 1.255 & 60 & 702 & - & - & - \\
\hline
 \end{tabular}
 \end{table*}
\begin{table*}
 \contcaption{A table continued from the previous one.}
 \label{tab:continued}
 \begin{tabular}{lccccccccccc}
 \hline
 Case No.& $M_{\ast}$ & Disk slope & Migration & $r_{\rm trans}$ & $r_{\rm ice}$ & $N_{\rm emb}$ & $N_{\rm plt}$& $N_{\rm p}$ &$M_{\rm pmax}$ & Accretion efficiency & Water delivery \\
&($M_{\odot}$)& & & (AU) & (AU) & & & & ($M_{\oplus}$) & & \\
\hline
%%G3-19& 0.6 & 2 & Yes & 0.337 & 1.255 & 60 & 702 & 4 & 15.335 & 98.01\% & - \\
%%G3-20& 0.6 & 2 & Yes & 0.337 & 1.255 & 60 & 702 & 4 & 16.78 & 96.72\% & - \\
G3-19& 0.6 & 3/2 & Yes & 0.337 & 1.255 & 60 & 702 & 3 & 11.595 & 85.61\% & - \\
G3-20& 0.6 & 3/2 & Yes & 0.337 & 1.255 & 60 & 702 & 3 & 10.630 & 88.18\% & - \\
G3-21& 0.6 & 2 & Yes & 0.337 & 1.255 & 60 & 702 & 4 & 15.335 & 93.16\% & - \\
G3-22& 0.6 & 2 & Yes & 0.337 & 1.255 & 60 & 702 & 4 & 16.78 & 93.02\% & - \\
\hline
\end{tabular}
\begin{tablenotes}
    \footnotesize
    \item[]Note: $N_{emb}$ and $N_plt$ represent the number of the embryos and planetesimals, respectively. $M_{pmax}$ is the maximum mass of planets formed.
    \end{tablenotes}
\end{table*}

\section*{Acknowledgements}
We thank the anonymous referee for constructive comments
and suggestions to improve the original manuscript.
This work is financially supported by the B-type Strategic Priority Programme of the Chinese  Academy of Sciences (Grant No. XDB41000000), the National Natural
Science Foundation of China (Grant Nos. 12033010, 11773081, 12111530175),
the Strategic Priority Research Program on Space  Science of the Chinese Academy of Sciences (Grant No. XDA15020800), the China Manned Space Project with NO. CMS-CSST-2021-A11 and CMS-CSST-2021-B09, Foundation of Minor Planets of the Purple Mountain Observatory, and Youth Innovation Promotion Association CAS.

%%%%%%%%%%%%%%%%%%%%%%%%%%%%%%%%%%%%%%%%%%%%%%%%%%
\section*{Data Availability}

The data underlying this article will be shared on reasonable request
to the corresponding author.



%\bibliographystyle{mnras}

\bsp    % typesetting comment
\label{lastpage}
\end{document}